\documentclass[superscriptaddress,floatfix,aps,pre,twocolumn,notitlepage,shortbibliography,nofootinbib]{revtex4-2}

\usepackage[utf8]{inputenc}
\usepackage{amsmath}
\usepackage{amssymb}
\usepackage{refcount}
\usepackage{siunitx}
\usepackage{graphicx}
\usepackage{hyperref}
\usepackage{natbib}
 \usepackage[capitalise]{cleveref}
\usepackage{microtype}
\usepackage{bm}
\usepackage{lipsum}
\usepackage[english]{babel}
\usepackage{color}
\usepackage{graphicx}
\usepackage{xcolor}
\usepackage{braket}

\begin{document}
\author{Haidar Al-Naseri}
\email{haidar.al-naseri@umu.se}
\affiliation{Department of Physics, Ume{\aa} University, SE--901 87 Ume{\aa}, Sweden}

\author{Gert Brodin}
\email{gert.brodin@umu.se}
\affiliation{Department of Physics, Ume{\aa} University, SE--901 87 Ume{\aa}, Sweden}
\title{Radiation reaction effects in relativistic plasmas - the electrostatic limit}

\begin{abstract}
We study the evolution of electrostatic plasma waves, using the relativistic Vlasov equation extended by the Landau-Lifshitz radiation reaction, accounting for the back-reaction due to the emission of single particle Larmor radiation. In particular, the Langmuir wave damping is calculated as a function of wavenumber, initial temperature, and initial electric field amplitude. Moreover, the background distribution function loses energy in the process, and we calculate the cooling rate as a function of initial temperature and initial wave amplitude. Finally, we investigate how the relative magnitude of wave damping and background cooling varies with the initial parameters. In particular, it is found that the relative contribution to the energy loss associated with background cooling decreases slowly with the initial wave amplitude.
\end{abstract}
 
\maketitle

\section{Introduction}
In the past decades, there has been a steady increase in the interest of intense field plasma physics. This has been driven in part by technological development (see,  Ref. \cite{High-intensity} for the current high-intensity laser world record, and Refs. \cite{projected1,projected2} for the projected performance of upcoming facilities), by experimental findings (e.g. Ref. \cite{experiments}), and by theoretical concerns (e.g.  Refs. \cite{QED-review1,QED-review2,QED-review3})] 

Electrons are nowadays accelerated to the strongly relativistic regime rather routinely, see e.g. Refs. \cite{QED-review1,QED-review2,QED-review3}). As is well known, strongly accelerated relativistic electrons emit Larmor radiation. Unless the emitted photons are very hard, this process will be well described by the relativistic generalization of Larmor's formula. As a consequence of this self-interaction of the electron with its own field, there is an effective recoil force on the electrons, that should be added to the Lorentz force of the external electromagnetic field. 

The nature of the recoil force, also known as radiation reaction, has been studied extensively for a long time, see e.g. Refs. \cite{Landau_Lif,Jackson}. In the classical regime, excluding the strongest field intensities, a key result is given by the Abraham-Lorentz-Dirac (LAD) equation. However, the LAD equation is famous for the unphysical runaway solutions. Treating the recoil force as a small perturbation to the Lorentz force, however, the unphysical solutions can be removed, in which case we are left with the Landau-Lifshitz (LL) equation for the radiation reaction. A tutorial review where many related aspects are covered is given in Ref. \cite{Burton} 

The Landau-Lifshitz equation has been solved exactly for some restricted field geometries, see e.g.  Refs. \cite{Bulanov}, for the one-particle system.
Numerically, the LL radiation reaction has been implemented in particle-in-cell (PIC) simulation by Ref. \cite{Silva,Wallin}. Moreover, a quantum generalization of LL radiation reaction has been implemented in PIC simulations by Refs. \cite{Silva_Q,Wallin}.
For a comparison between classical and quantum radiation reaction in PIC simulations, see Ref. \cite{Wallin}. 

For kinetic theories of a many-particle system, Hakim et al. were the first to derive radiation reaction correction to the Vlasov equation \cite{Hakim}. Radiation reaction corrected Vlasov equations have also been derived in \cite{Kunze,Burton,Elskens}. The focus of the above-mentioned works was to derive kinetic evolution equations of the system, rather than analyzing the resulting dynamics. A kinetic study of Landau damping influenced by radiation reaction effects was presented in Ref. \cite{Burton-2}. Hydrodynamic models of relativistic plasmas including radiation reaction have also been studied, see e.g. Refs. \cite{Mahajan1,Mahajan2,Mahajan3,Dalakishvili}.

In this work, we focus on the effects of radiation reaction in the presence of electrostatic waves.  For this purpose, we add the Landau-Lifshitz expression for the radiation reaction to the Lorentz force in the relativistic Vlasov equation. The influence of radiation reaction on Langmuir waves is then studied. As expected, the radiation reaction induces wave damping, and the damping is computed numerically as a function wave-number, initial temperature, and initial electric field amplitude. In particular, it is found that the normalized energy loss rate (with respect to the initial energy) decays with a factor $\sim 2/3$ with increasing amplitude for a low or modest temperature. This decrease is found to be a direct consequence of the transition from a sinusoidal wave profile (in the low amplitude regime) to a saw-tooth profile (in the strongly relativistic limit). 

Moreover, it is found that wave damping occurs simultaneously as the background electron distribution loses kinetic energy, i.e. the radiation reaction induces electron cooling. The magnitude of the cooling is studied as a function of the initial temperature and the initial electric field. Generally, the Larmor radiation takes energy from two sources, wave damping, and the background electron distribution. The relative magnitude of these contributions is investigated, and it is found that electron cooling dominates for high background temperature and strong electric fields, whereas the opposite ordering applies in the low-temperature weak field regime.

The organization of the paper is as follows: In section II, the basic equations are presented, simplifications for the present case of electrostatic waves are made in II A, and results for the low-temperature regime are derived in II B. Section III are devoted to numerical studies concerning the dependence of wave damping on the wave number (III A), the scaling of the cooling rate (III B) with amplitude and temperature, and the relative magnitude of cooling and wave damping (III C). Finally, our results are discussed in section IV.    

\section{Basic equations}

For sufficiently strong electromagnetic fields, the relativistic Vlasov equation for electrons needs to be updated \cite{QED-review1,QED-review2,QED-review3, E-schwinger}. For field strengths well below the Schwinger critical field, electron-positron pair production due to the Schwinger mechanism can be neglected. However, photon emission by single electrons due to nonlinear Compton scattering may become significant in case the product $\chi a_0^2$ is not too small. Here we have introduced the quantum nonlinearity parameter $\chi$ \cite{QED-review1} covariantly written as 
\begin{equation*}
    \chi=\frac{1}{E_{cr}c } \sqrt{F^{\mu\nu}u_{\nu}F_{\mu\sigma}u^{\sigma}}
\end{equation*}
which is 
typically much smaller than unity. Moreover, we have introduced the "laser strength" $a_0=eE/m\omega$ (roughly the relativistic gamma factor due to electron quiver velocity for large electric fields $E$, in which case $a_0$ is larger than unity). Here $\omega$ is the wave frequency, $F^{\mu\nu}$ the electromagnetic field tensor, $u^\mu$ the four-velocity, $c$ the speed of light in vacuum, $E_{cr}=m^2c^3/e\hbar$ the Schwinger critical field, $m$ and $e$ being the electron mass and charge respectively, and, finally, $\hbar$ is the reduced Planck constant. Note, that  $e$ is the negative electron charge, as opposed to the positive elementary charge.

Many works (see the recent reviews \cite{QED-review1,QED-review2,QED-review3} for long lists of papers) have studied nonlinear Compton scattering for a small electron number density, such that the driving electromagnetic fields can be taken as solutions to Maxwell's equations in vacuum, in which case the properties of the emission spectra are of main concern. However, in case the electron number density is higher, the plasma dynamic for strong fields is far from trivial. Importantly, in case the energy radiated by the electrons is not very small, the electron equation of motion (given by the Lorentz force in the external field) needs to be corrected by the radiation reaction force \cite{Wallin, Silva, Silva_Q}.  

While there is ongoing research on how to extend the classical expression by Landau-Lifshitz (LL), to include QED-physics as well as higher-order classical effects (see e.g. Refs. \cite{QED-review1,QED-review2,QED-review3}), in the regime where radiation reaction can be considered a small perturbation, the LL-expression is well established \cite{QED-review1,QED-review2,QED-review3}. Since we will here consider this particular regime, we adopt the LL-expression given by Refs. \cite{Landau_Lif,Jackson,Bulanov}

\begin{multline}
\label{RR}
    \mathbf{F}=\frac{2e^3\epsilon}{3m^2c^5}
    \Big(\partial_t+ \frac{\mathbf{p}}{\epsilon}\cdot\nabla_x \Big)
    \bigg[ \mathbf{E} + \frac{c\mathbf{p}}{\epsilon}\times \mathbf{B} \bigg] \\
    +\frac{2e^4}{3m^2c^4}
    \bigg[ \mathbf{E}\times \mathbf{B}+  \mathbf{B}\times \Big(\mathbf{B}\times \frac{c\mathbf{p}}{ \epsilon}\Big)
    + \mathbf{E}\Big(\frac{c\mathbf{p}}{ \epsilon}\cdot\mathbf{E}\Big)
    \bigg]\\
    -\frac{2e^4 \epsilon}{3c^7m^4}\mathbf{p}
    \bigg[\Big(\mathbf{E}+ \frac{c\mathbf{p}}{\epsilon}\times \mathbf{B}\Big)^2
    -(\frac{c\mathbf{p}}{\epsilon}\cdot \mathbf{E})^2
    \bigg]
\end{multline}
where we have introduced the particle energy $\epsilon=mc^2\sqrt{1+p^2/m^2c^2}$. Given \cref{RR}, the relativistic Vlasov equation, with the radiation reaction $\mathbf{F}$ as a correction to the Lorentz-force, can be written 
\begin{equation}
   \Big[ \frac{\partial}{\partial t} +\frac{\mathbf{p}}{\epsilon}\cdot \nabla \Big]f + e
   \Big(\mathbf{E}+ \frac{c\mathbf{p}}{\epsilon} \times \mathbf{B}\Big)\cdot \nabla_pf + \nabla_p\cdot(\mathbf{F}f)
\end{equation}
Note here that particle conservation demands the correction term to be written $\nabla_p\cdot(\mathbf{F}f)$ rather than  $\mathbf{F}\cdot\nabla_pf$, since, contrary to non-dissipative forces such as the Lorentz force, $\nabla_p\cdot\mathbf{F}\neq 0$.  It should be stressed that after introducing the radiation reaction in the Vlasov equation, the Maxwell-Vlasov system will not be energy-conserving anymore, as the macroscopic current $-e\int({\bf p}f/\epsilon d^3p$ will not resolve the motion of individual particles, leading to the emission of high-frequency Larmor radiation, constituting the missing piece in the energy balance. In this work, we will ignore the effects of the generated high-frequency radiation on the collective dynamics, apart from what is already captured by the radiation reaction. The basic assumption is that any additional influence is a higher-order effect that is not crucial until the radiation reaction becomes comparable in magnitude to the Lorentz force. 
\subsection{Kinetic theory for electrostatic fields}

From now on we will consider the one-dimensional electrostatic limit with ${\bf E}=E(z,t){\hat {\bf z}}$, in which case  the radiation reaction force reduces to
\begin{equation}
    \mathbf{F}=\frac{2e^3 \epsilon}{3m^2c^5}
    \bigg[ \Big(\frac{\partial E}{\partial t} + \frac{p_z}{\epsilon } \frac{\partial E}{\partial z}\Big)\mathbf{e}_z
    + \frac{c^2eE^2p_z}{\epsilon^2}\mathbf{e}_z
    -\frac{eE^2\epsilon_{\bot}^2}{m^2c^2\epsilon^2}\mathbf{p}
    \bigg]
    \label{RR1}
\end{equation}
As a consequence, in the electrostatic 1D limit, the relativistic Vlasov equation including radiation reaction is given by
\begin{multline}
\label{Rel_vlasov_estat}
       \Big[ \frac{\partial}{\partial t} +\frac{p_z}{\epsilon} \frac{\partial}{\partial z} \Big]f 
       + \frac{e}{m}
   E\frac{\partial f}{\partial p_z}
      -\frac{1}{p_{\bot}}\frac{\partial }{\partial p_{\bot}} \Bigg(\frac{2 e^4E^2\epsilon_{\bot}^2 p_{\bot}^2}{3m^4c^5\epsilon^2} f\Bigg)\\
   +  
   \frac{2e^3 }{3m^2c^5} \frac{\partial}{\partial p_z}
    \bigg[ \Big(\epsilon \frac{\partial E}{\partial t} + p_z \frac{\partial E}{\partial z}\Big) f
    + \frac{eE^2p_zp_{\bot}^2}{\epsilon m^2} f \bigg]=0
\end{multline}
To prepare for numerical calculations, we introduce the following normalized variables :
\begin{align}
\label{Normalization}
    t_n&=\omega_p t \notag \\
    z_n&=\frac{\omega_p z}{c}\notag  \\
    p_n&= \frac{p}{mc} \notag  \\
    \epsilon_n&=\frac{\epsilon}{mc^2}\\
    f_n&= \frac{m^3c^3}{n_0}f \notag  \\
    E_n&= \frac{eE}{mc\omega_p} \notag 
\end{align}
where $\omega_p\equiv\sqrt{n_0e^2/\epsilon_0 m}$, and $n_0$ is the unperturbed electron number density. Applying the normalization of \cref{Normalization} in \cref{Rel_vlasov_estat} we get
\begin{multline}
\label{Normalized-estat}
     \Big(\frac{\partial}{\partial t_n}+´\frac{p_{zn}}{\epsilon_n}\frac{\partial}{\partial z_n}+ E_n\frac{\partial}{\partial p_{zn}}\Big)f_n \\
     - \frac{1}{p_{\bot n}} \frac{\partial }{\partial p_{\bot n}} \bigg( \frac{2 \delta\epsilon_{\bot n}^2p_{\bot n}^2E_n^2}{3\epsilon_n}f_n\bigg)\\
         + \frac{2\delta }{3} \frac{\partial }{\partial p_{zn}} \bigg[ \Big(\epsilon_n \frac{\partial E_n}{\partial t_n} 
    + p_{zn}\frac{\partial E_n}{\partial z_n}
     - \frac{E_n^2p_{zn} p_{\bot n}^2 }{\epsilon_n}\Big)f_n     \bigg]
     =0
\end{multline}
where $\delta= r_e \omega_p/c$ and $r_e$ is the classical electron radius, that is 
    $r_e=e^2/mc^2$
in the cgs unit system. We note that except for extremely high-density plasmas (like, for example, the central parts of neutron stars), $\delta\ll1$ applies, which will be used throughout the manuscript.  
  Finally, the radiation reaction corrected Vlasov \cref{Normalized-estat}, is complemented by Ampère's law to obtain a closed system.
\begin{equation}
\label{Ampers_law}
    \frac{\partial E_n}{\partial t_n}= -\int d^3p_n\frac{p_{zn}}{\epsilon_n}f_n
\end{equation}
For notational convenience, in what follows, we will drop the subscript $n$ on the normalized variables. 
The total energy $W_{tot}$ in the system is the sum of the electrostatic energy and the kinetic and rest mass energy, that is 
\begin{equation}
\label{Tot_energy}
    W_{tot}=\frac{E^2}{2}+ \int d^3p\, \epsilon f
\end{equation}
Since we have energy loss in the system due to the Larmor high-frequency radiation, the total energy is not conserved.  Taking the time-derivative of \cref{Tot_energy} and using \cref{Normalized-estat} we get
\begin{equation}
\label{Econs}
    \frac{\partial W_{tot}}{\partial t}+\frac{\partial J}{\partial z}= \frac{2\delta}{3} \int d^3p \bigg(p_{z} \bigg(\frac{\partial E}{\partial t}+
    \frac{p_z}{\epsilon}\frac{\partial E}{\partial z}\bigg)- E^2p_{\bot }^2\bigg) f
\end{equation}
where the energy flux $J$ is
\begin{align}
     J= \int d^3p\, p_{z}f
\end{align}
We note that the right-hand side of \cref{Econs} is negative definite, as should be expected since this term represents the energy loss due to short-scale electromagnetic degrees of freedom not resolved by the macroscopic current computed in \cref{Ampers_law}. The negative sign of this term can be formally proven, noting that the time derivative of the electric field can be moved outside the momentum integral. Then the use of \cref{Ampers_law} is enough to assure the sign of the first term of the right-hand side of \cref{Econs}, and the second term is explicitly negative to start with. 

For the case $\delta \ll 1$, which will be of main interest, the energy loss given by the right-hand side of \cref{Econs} can be viewed as a small perturbation. In this regime, the wave damping rate will be linear in $\delta$.   As we will see in the next section, this means that \cref{Econs} can be of help when evaluating the damping rate of Langmuir waves. 

\subsection{The low-temperature limit}

Before we take on the more general case,  it is instructive to first consider the low-temperature limit. For a sufficiently low temperature, kinetic effects can be neglected and the evolution equations can be simplified. For this purpose, we define the following moments over momentum space
\begin{align*}
    n(z,t)&=\int  d^3p f\\
    P(z,t)&=\int d^3p \frac{p_zf }{  n(z,t)}
\end{align*}
and assume that the spread in momentum (temperature) is low enough such that we can make the approximation
\begin{equation*}
    h(P(z,t))\approx \int  d^3p\frac{h(p_z)f}{ n(z,t)}
\end{equation*}
 where $h(p_z)$ is an arbitrary function of $p_z$. 
Next, integrating \cref{Normalized-estat} over momentum, we immediately obtain the continuity equation
\begin{equation}
\label{cold1}
    \frac{\partial n}{\partial t}=-\frac{\partial}{\partial z}\bigg(\frac{P n}{\sqrt{1+P^2}}\bigg).
\end{equation} 
Furthermore, we multiply \cref{Normalized-estat} with $p_z$, integrate over momentum space, and apply the low-temperature approximation. Rewriting the derivatives on the electric field using Maxwell's equations, we get the cold momentum equation with radiation reaction included, which can be written 
\begin{eqnarray}
\label{Momentu}
\frac{\partial P}{\partial t}=E- \frac{\partial}{\partial z}\sqrt{1+P^2} -\frac{2}{3}\delta P 
\end{eqnarray}
Finally, applying the low-temperature approximation to Ampere's law, we obtain 
\begin{equation}
\label{cold-ampere}
         \frac{\partial E}{\partial t}=-\frac{P n}{\sqrt{1+P^{2}}} 
\end{equation}
The energy conservation law for the system \cref{cold1}-\cref{cold-ampere} can be written as

\begin{equation}
\label{energy-cold-eq}
\frac{\partial W}{\partial t} +\frac{\partial J}{\partial z}
=-\delta \frac{2}{3}
 \frac{P^2n}{\sqrt{1+P^{2}}}
\end{equation}
where, in the cold limit, the energy density and energy flux are given by
\begin{equation}
\label{Energy_cold}
    W= \frac{1}{2}E^{2}+n(\sqrt{1+P^{2}}-1)
\end{equation}
and $J= nP$, respectively. We note that the third loss terms in \cref{Econs} vanishes in the cold limit, and the surviving loss term in \cref{energy-cold-eq} is a combination of the two previous ones in \cref{Econs}.  

\begin{figure}
    \centering
    \includegraphics[width=9cm,height=8cm]{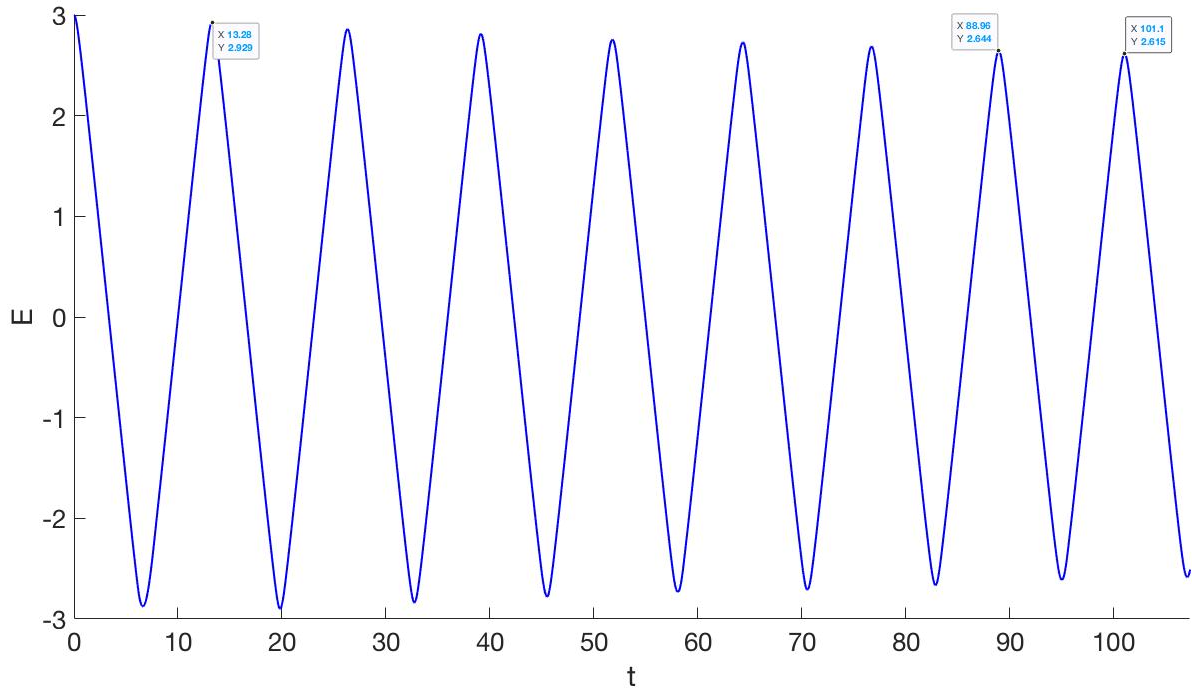}
    \caption{The electric field $E$ over time for  $\delta=0.005$ using the low-temperature limit.}
    \label{En_vs_tn}
\end{figure}

Before studying the dynamics of Langmuir waves in more detail in the next section, let us first illustrate some features for the simple case of no spatial dependence, i.e. we drop all terms with spatial derivatives in Eqs. (\ref{cold1})-(\ref{cold-ampere}). Starting with the initial values  $E(t=0)=E_0=3$, $P(t=0)=0$, $n(t=0)=1$, and letting $\delta=0.005$, the system (\cref{cold1})-(\cref{cold-ampere}) is solved numerically in the homogeneous limit. The result for the electric field is displayed in \cref{En_vs_tn}. The first thing to note is the saw-tooth profile of the electric field, which is a result of the relativistic amplitude. With the peak electric field of the order $E\sim 3$, the peak relativistic gamma factors are $\gamma\sim 9$ (for $E>1$, the peak value of the momentum scale as $P\sim E^2$), and for most of the plasma oscillation cycle the electron velocities are close to the speed of light, in which case the current is more or less constant (since the number density is conserved), and thus  the sawtooth-profile follows from Ampere's law. Secondly, we note that the loss rate is a few percent per oscillation cycle, which is well in accordance with \cref{energy-cold-eq}, taking into account that the loss rate of $(2/3)\delta$ is magnified by a factor $\langle  P^2/\sqrt{1+P^2}\rangle\sim\langle\gamma\rangle\sim 4-5$, where the last estimate refers to the data of \cref{En_vs_tn}. Here $\langle ...\rangle$ denotes averaging over an oscillation cycle. Thirdly, by comparing the time period for the first and last oscillation cycle, we note that the wave frequency is increasing. This is a natural consequence of the decreased wave amplitude, as the relativistic gamma factor $\gamma$ averaged over a wave cycle decreases with the wave amplitude, and the wave frequency scale as $\omega^2 \propto 1/\langle\gamma\rangle$.

Finally, we note that the energy loss rate per wave cycle decreases only slightly from cycle to cycle (compare the peak-to-peak changes of the electric field for the first and last wave cycle), as can be expected for a small value of $\delta$. This property will be used in the next section to solve the governing equations, \cref{Normalized-estat} and \cref{Ampers_law}, perturbatively.


\section{Numerical results}
\begin{figure}
    \centering
    \includegraphics[width=8 cm, height=7cm]{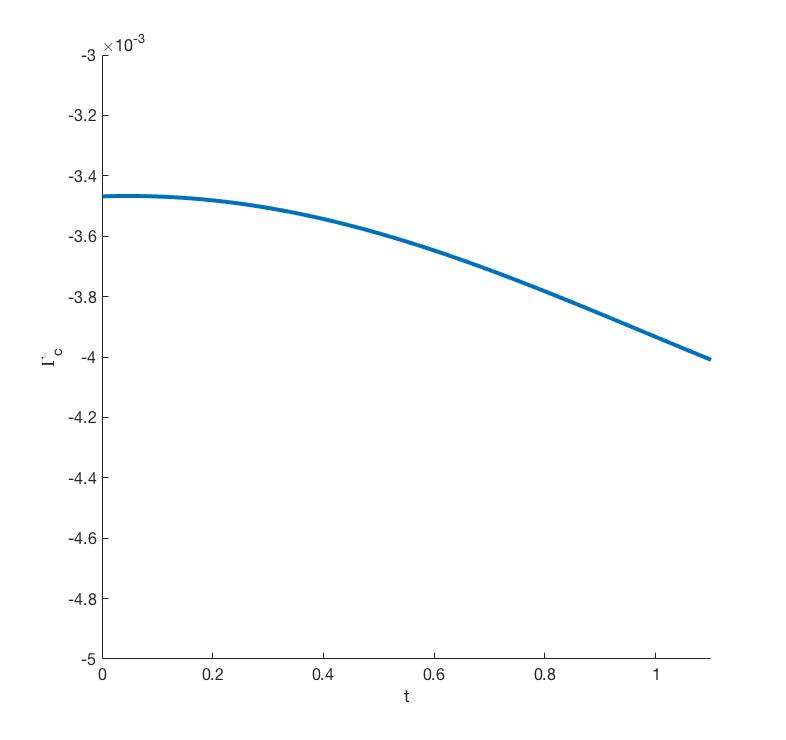}
    \caption{The evolution of the energy loss rate in the cold limit (given by spatial integration over the right-hand side of Eq. \ref{energy-cold-eq}) as a function of time, for $E_0=1$, $k=1$ and $\delta=0.02$.}
\label{W_vs_t}
\end{figure}
\begin{figure}
    \centering
    \includegraphics[width=0.5\textwidth]{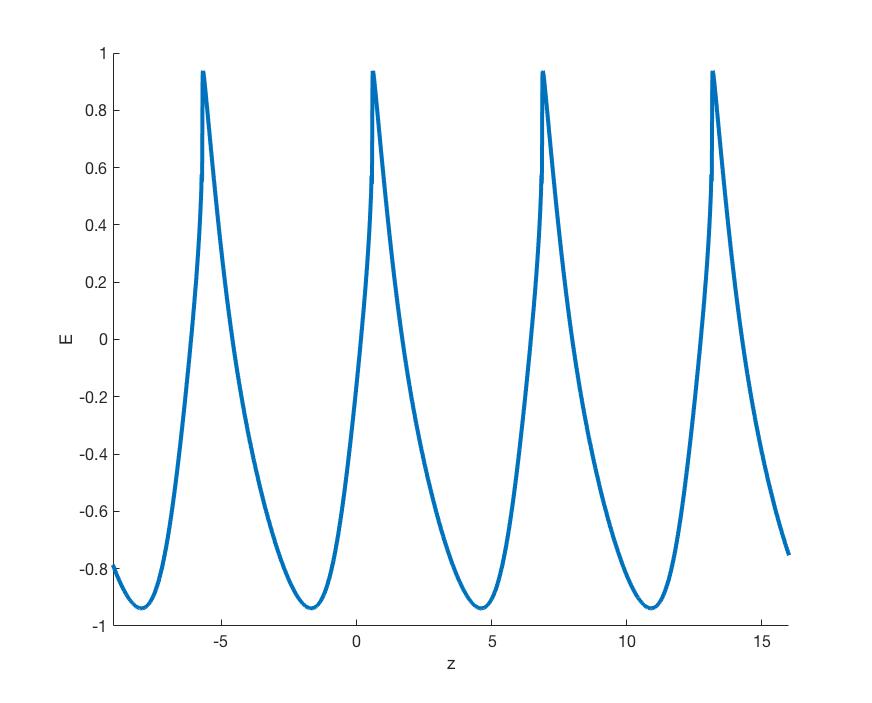}
    \caption{The spatial dependence of the electric field just before wave-breaking occurs, at $t=1.1$.  The used parameters  were $E_0=1$ and $k=1$. } 
    \label{Wave-breaking}
\end{figure}

In this section, we will perform a systematic numerical analysis to study how electrostatic plasma waves are affected by radiation reaction. This will include a dependence of the wave damping on wave number, temperature, and on the initial electric field amplitude. Moreover, we will investigate the cooling of the background electron distribution, which is a process induced by radiation reaction that accompanies wave damping.
\subsection{Dependence of wave damping on the wave-number}

Before analyzing the case with a finite temperature, we first would like to study the energy loss of Langmuir waves for a cold plasma, in particular the dependence of the loss rate on the wave number, based on the cold governing Eqs. (\ref{cold1})-(\ref{cold-ampere}). 
These equations have been solved numerically using a modified version of the Lax-Wendroff method \cite{Lax-Wenderoff}, with the following initial conditions
\begin{align*}
   E(t=0)&=E_0\cos(kz)  \\
   n(t=0)&=1-kE_0\sin(kz)\\
   P(t=0)&=-E_0\sin(kz).
\end{align*}
 The damping rate can be computed, keeping $kE_0>1$, to assure that the initial density is always positive. 
Next, we define the spatially integrated energy loss rate $\Gamma_c$ as 
\begin{equation}
\label{damping-rate}
\Gamma_c=\frac{1}{\int W dz}\int \frac{dW}{dt} dz        
\end{equation}
Here  the integral is carried out over all space, or alternatively, for the case of a spatially periodic function, we can limit the spatial integration to a single wavelength.
\begin{figure}
    \centering
    \includegraphics[width=0.5\textwidth]{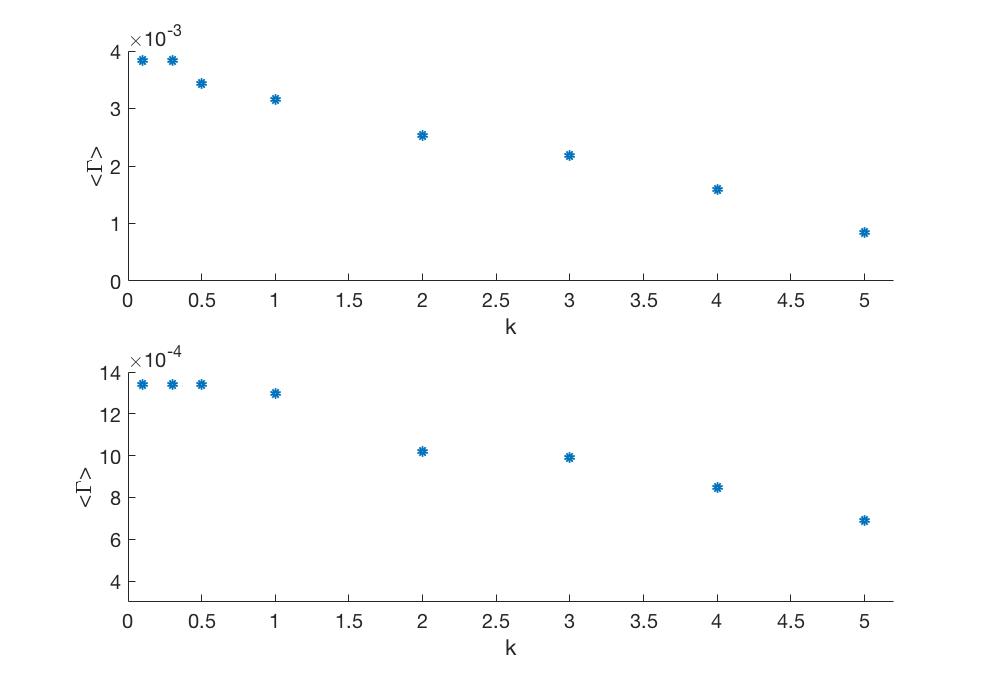}
    \caption{The energy loss rate $\Gamma$ of a Langmuir wave for $\delta=0.02$ as a function of wave number for two values of the initial electric field amplitude,  for $E_0=1$ (upper panel) and for $E_0=0.5$ (lower panel).}
\label{Gamma_vs_k}
\end{figure}
The energy loss rate defined in \cref{damping-rate} is the instantaneous one, which should be expected to vary in the nonlinear regime. However, the variation with time is typically rather modest, as displayed in \cref{W_vs_t} for the case of $E_0=1$, $k=1$, and $\delta=0.02$. We note that the energy loss rate is decreasing slowly, indicating that particles are pushed away from the regions of a higher electric field by the ponderomotive force, decreasing the energy loss rate (cf. \cref{energy-cold-eq}). However, for the present numerical runs, we note that typically the difference between the minimum and the maximum loss rate is not larger than $\sim 10\%$, even for nonlinear initial conditions. 

Next, we should be aware that systems that are nonlinear, cold, and spatially varying (nonzero $k$), tend to undergo wave-breaking eventually.  When wave-breaking occurs, the single-valued fluid momentum becomes two-valued, due to one electron fluid element overtaking another. At this point, the fluid description ceases to be applicable. While a cold electron fluid is most susceptible to wave-breaking, the process can occur also in warm fluid theory, see e.g. Ref. \cite{wavebreaking}. In the present subsection, where we limit ourselves to cold fluid theory, we solve Eqs. (\ref{cold1})-(\ref{cold-ampere}), in order to evaluate the energy loss rate up to the point where wave-breaking sets in, and the fluid limit ceases to be applicable. In \cref{Wave-breaking}, the electric field profile is shown right before wave-breaking sets in - note the almost infinite spatial derivative indicating wave-breaking is about to occur. Here the initial conditions are a harmonic spatial profile with the same parameters as in figure 1.  

Next, we study the energy loss rate $\langle \Gamma \rangle$ as a function of wavenumber $k$. Here $\langle \Gamma \rangle$ is computed as the average loss rate from $t=0$ up to the wave-breaking time $T_{W}$, i.e. 
\begin{equation*}
    \langle \Gamma \rangle=\frac{1}{T_w} \int_0^{T_{W}} \Gamma(t)dt.
\end{equation*}

In \cref{Gamma_vs_k} we see the energy loss rate for different wavenumbers for initial electric field $E_0 =1$ (first panel) and for $E=0.5$ (second panel). The general feature is a decline of the loss rate with wavenumber in both cases. The reason for the decline is that the loss rate is proportional to $P^2n/\sqrt{1+P^2}$ (see \cref{energy-cold-eq}) and that for higher spatial gradients, the ponderomotive force increases and can push particles to regions of a lower electric field more effectively, limiting the average momentum  and thereby the loss rate. Furthermore, we see that the loss rate is considerably larger for a stronger electric field, as should be expected from the nonlinear dependence of the loss rate on momentum. Here it is important to note that $P$ scales as $E^2$ for larger fields well beyond the linear regime.  

\subsection{Cooling due to radiation reaction}
Generally speaking, radiation reaction provides a mechanism for the transfer of (electron) kinetic energy to high-frequency EM radiation. However, except for the cold case $T=0$, not all of the electron kinetic electron energy is associated with the wave motion. Most previous studies (as seen e.g. in Ref. \cite{Mahajan3}) have found a cooling of the background distribution due to radiation reaction. However, as shown by Ref. \cite{Di-Piazza} in the regime of quantum radiation reaction, also heating of the background distribution is possible. To study the energy transfer in more detail, in this sub-section we focus on the cooling of the background distribution due to classical radiation reaction. 

Firstly, we need to consider some preliminaries, in order to set up a perturbative approach making use of $\delta\ll 1$, applicable when the radiation reaction can be used as a small correction to the external Lorentz force. Considering the homogeneous limit, $k=0$, we solve the relativistic Vlasov by making a formal change of variables. For this purpose, we introduce the canonical momentum
\begin{equation}
    q=p_z-A(t)
\end{equation}
where $A(t)$ is the normalized vector potential, in which case the relativistic Vlasov equation (dropping radiation reaction) \cref{Normalized-estat} becomes
\begin{equation}
\label{Vlasov_qtranform}
    \frac{\partial }{\partial t}f(q,p_{\bot},t)=0
\end{equation}
together with Ampére's law
\begin{equation}
\label{Amper'slaw2}
    \frac{\partial E (t)}{\partial t}=A(t)\int d^3p \frac{1}{\epsilon} f(q,p_{\bot},t=0)
\end{equation}
where $\epsilon=\sqrt{1+p_{\perp}^2+(q-A(t))^2}$. Note that a term has been dropped in \cref{Amper'slaw2} after integration, due to $f$ being an even function of $q$.
Solving \cref{Amper'slaw2} (together with $E=-\partial A/\partial t$) for $E(t=0)=E_0$, we can then use the resulting temporal profile, $E(t)$, as our input in a perturbative scheme. As our initial background distribution (which is preserved to zero:th order in canonical momentum coordinates, see \cref{Vlasov_qtranform}) we will consider a Maxwell-J{\"u}ttner distribution $f_0$, i.e. we let
\begin{equation}
\label{background}
    f_0= \frac{1}{\int e^{-\sqrt{1+p_{\perp}^2+q^2}/E_{th}}p_{\perp}dp_{\perp}dq} \, e^{-\sqrt{1+p^2}/E_{th} }
\end{equation}
where the thermal energy is $E_{th}=\sqrt{1+p_{th}^2}-1$ and
$p_{th}$ is the normalized thermal momentum. 

Strictly speaking, even if we begin with a thermodynamic background distribution as in \cref{background}, as soon as the generation of high-frequency radiation starts, we do not have a well-defined temperature in the thermodynamic sense. Still, counting all of the kinetic energy not associated with the oscillatory net drift as thermal, we can nevertheless define an effective temperature associated with the evolving background distribution, in order to produce a quantitative description. 

At $t=0$, we can thus define the initial (normalized) temperature $T_0$ as 
\begin{equation}
    T_0=\frac{2}{3}\left[ \int \left(\varepsilon -1\right)fp_{\perp}dp_{\perp}dq
\right] 
\end{equation}
where the initial drift momentum is $P=\int p f p_{\perp}dp_{\perp}dq=0$ (with the normalized density conserved, i.e. $n=1$). For the moment, we still ignore the effect of radiation reaction and assume that the dynamic is governed by the Vlasov equation, such that \cref{Vlasov_qtranform} applies. Defining the temperature as a function of time according to   
\begin{equation}
    T(t)=\frac{2}{3}\left[ \int \left(\sqrt{1+(p_z-A(t))^2+p_{\perp}^2} -1\right)fp_{\perp}dp_{\perp}dq
\right], 
\end{equation}
since $p_z-A(t)=q$, we see from \cref{Vlasov_qtranform} that this implies temperature conservation, $dT/dt=0$. 
\begin{figure}
    \centering
    \includegraphics[width=9 cm, height=7 cm ]{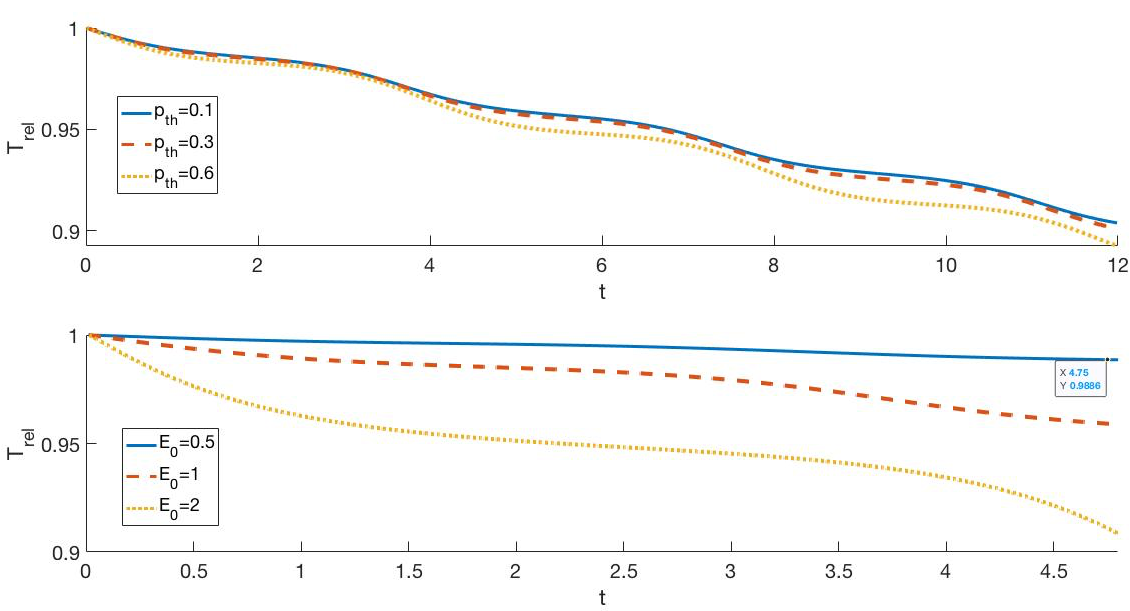}
    \caption{ $T_{rel}$ as a function of time $t$. In the first panel we have $E_0=1$, $\delta=0.01$ and $p_{th}=({0.1,0.3,0.5})$. In the second panel we have $p_{th}=0.3$, $\delta=0.01$ and $E_0=({0.5,1,2})$ .}
    \label{Cooling}
\end{figure}

Next, we turn to the case with radiation reaction included. Assuming the added term to be a small correction, we let $f=f_v+\delta f$, where $f_v$ is a solution to the unperturbed Vlasov equation. We only follow the evolution as long as $\delta f \ll f_v$, that is, we limit ourselves to the initial cooling phase. Under these conditions, applying \cref{Normalized-estat} perturbatively, we can define $T=T_{0}+\delta T$, where $T_0$ is the (constant) initial temperature,  and the temperature change $\delta T\equiv(2/3)\delta W_c)$ (where $\delta W_c$ is the change of the background kinetic energy) is given by 
\begin{equation}
\label{temp3}
\delta T=\frac{2}{3}\int \left[ \sqrt{1+q^{2}+p_{\bot }^{2}}-1\right] \delta fp_{\bot
}dp_{\bot }dq 
\end{equation}%
together with%
\begin{equation}
\delta f=\int_{0}^{t}-\nabla _{\tilde{p}}\cdot \Big(\mathbf{F}%
_{rad} \,f_{v}(q,p_{\bot },t=0)\Big)dt^{\prime } 
\label{cooling-exp}
\end{equation}
Inserting the expression for \cref{cooling-exp} into \cref{temp3}, after a partial integration we can derive an expression for the rate of change of the effective temperature
\begin{equation}
\label{temp4}
\delta W_c=\int_0^t\frac{3}{2}\frac{dT}{dt'}dt'= \int_0^t\frac{dW_c}{dt'}dt'
\end{equation}
where the rate of change of the background kinetic energy is
\begin{multline}
\label{temp5}
\frac{dW_c}{dt}=\\
\frac{2\delta}{3} \int p_{\perp}dp_{\perp}dq  \bigg(q\frac{\partial E}{\partial t}- \frac{(1+p_{\perp}^2+q^2-qA)}{\epsilon^2}E^2p_{\bot }^2\bigg)\frac{\epsilon f_v}{\epsilon_q}
\end{multline}
with $\epsilon=\sqrt{1+p_{\perp}^2+(q-A)^2}$ as before, and we have introduced the notation $\epsilon_q=\sqrt{1+p_{\perp}^2+q^2}$. 

\begin{figure}
    \centering
    \includegraphics[width=0.5\textwidth ]{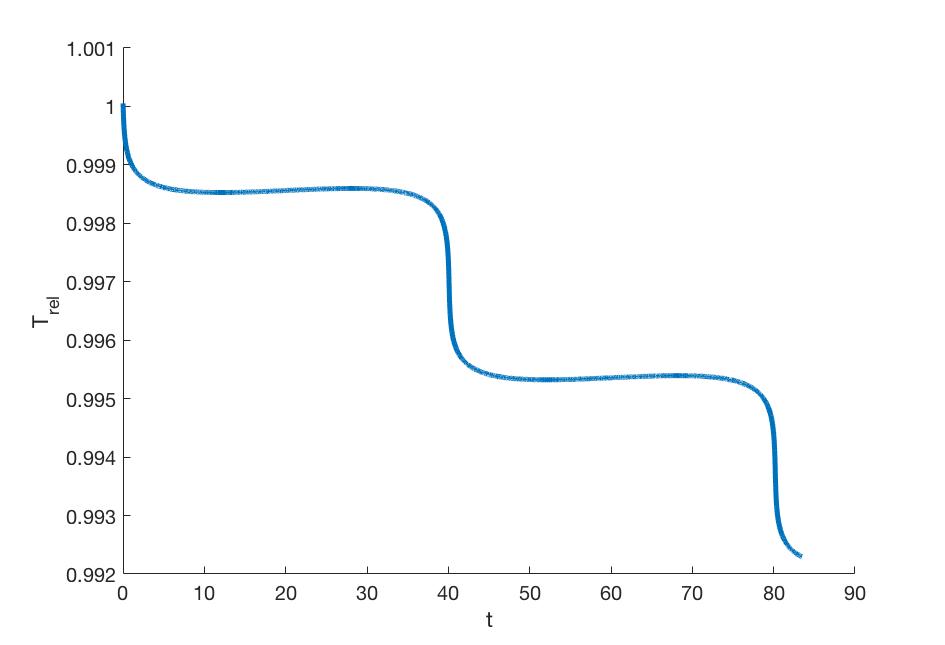}
    \caption{$T_{rel}$ for $\delta=10^{-5}$, $p_{th}=0.3$ .and $E=20$.}
    \label{fig:my_label}
\end{figure}

 In the upper panel of \cref{Cooling} the evolution of the normalized temperature $T_{rel}= (T_0+\delta T)/T_0$ is displayed for initial electric field $E=1$ and $\delta=0.01$ for different initial temperatures. We can see that the relative temperature decrease is only somewhat stronger for a higher temperature. While the relative difference in the cooling rate due to the difference in initial thermal energy is fairly modest, in absolute terms, naturally, the cooling is much more pronounced for a higher initial temperature.  

 In the lower panel of \cref{Cooling}, the evolution of the normalized temperature is shown for different values of the initial electric field, $E_0$, for $p_{th}=0.3$. As is obvious, the cooling rate shows a very strong dependence on the initial electric field. Roughly speaking, the temperature loss rate is proportional to $E_0^2$, as can be expected from \cref{temp5}.

Up to now, we have mostly used values of $\delta$ of the order $\delta\sim 10^{-2}$. While $\delta\ll 1$ still applies, such values correspond to very high densities, not common in a laboratory plasma context. Since we have been mainly interested in properties independent of $\delta$, e.g. the scaling of the energy transfer with temperature and initial amplitude, picking a somewhat larger value of $\delta$ is not necessarily a problem. However, it is of some interest to note that cooling due to radiation reaction for a plasma with a density slightly above solid density, $\delta=10^{-5}$, is not necessarily slow, provided the initial electric field amplitude is large. In \cref{fig:my_label} we follow the temperature evolution for roughly a plasma period, for $\delta=10^{-5}$, $E_0=20$, and $p_{th}=0.3$. The steep temperature drops occur two times per cycle when the absolute value of the electric field is close to its maximum. As can be seen, for the given initial data, the temperature drops roughly $1\%$ during a plasma period, i.e. after a modest number of oscillation cycles, $\sim70$, the temperature will have dropped roughly a factor of two. 

 A major question regarding cooling, not yet addressed, is the relative magnitude of the thermal energy drop in relation to the wave-damping energy loss. We will wait to consider this particular issue until we have studied wave damping in more detail.

\subsection{Wave damping in the homogeneous limit}

When the temperature is low, the energy loss induced by radiation reaction comes mainly in the form of wave damping, rather than cooling of the background distribution. Generally, however, the effects of cooling and wave-damping can be comparable in magnitude, and we need to separate the different contributions for a detailed description. For a small $\delta$, such that a perturbative approach is applicable, we can combine the total energy loss rate given by the right-hand side of \cref{Econs} with the cooling expression \cref{temp5} to identify the energy loss that corresponds to wave damping. Since the energy loss rate tends to vary during an oscillation cycle, naturally the loss rate should be evaluated during a full oscillation cycle, i.e. during $0\leq t\leq T_p$, where $T_p$ is the period time of the plasma oscillation. Using $\delta W_{tot}=\delta W_w + \delta{W_c}$ (where $\delta W_{tot}$ is the total energy loss and $\delta W_{w}$ the wave energy loss), with 
\begin{equation}
\label{Wloss}
\delta W_{tot} =     \int_0^{T_p}\frac{2\delta}{3} \int p_{\perp}dp_{\perp}dq \bigg(p_z \frac{\partial E}{\partial t}- E^2p_{\bot}^2\bigg) f_v dt
\end{equation}
and $\delta{W_c}=\delta{W_c}(t=T_p)$ computed from \cref{temp4} as in the previous sub-section, the wave damping rate can be expressed as   

\begin{equation}
    \Gamma_H= \frac{\delta W_{tot}-\delta W_c}{W_0 T_p}
    \label{Gamma_H}
\end{equation}
Here $W_0=E_0^2/2$ is the initial wave energy, and $f_v$ denotes the solution to \cref{Normalized-estat}, where the radiation reaction is dropped, in agreement with the perturbation scheme. 

\begin{figure}
    \centering
    \includegraphics[width=9cm,height=10cm]{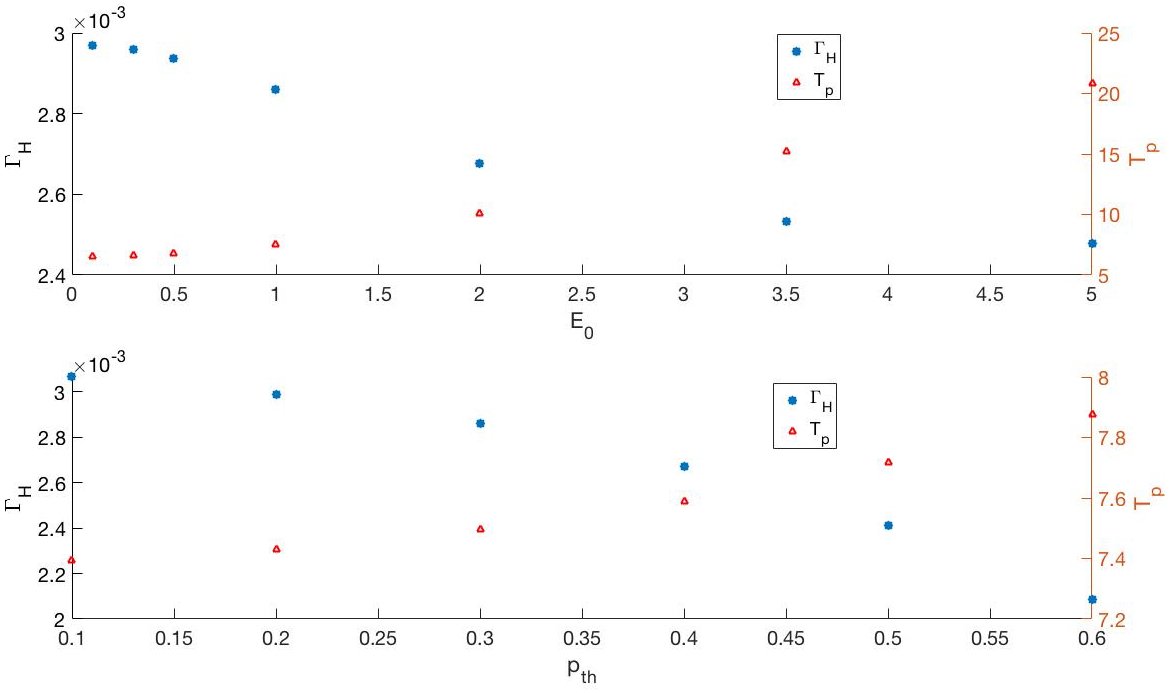}
    \caption{The dots show the damping $\Gamma_H$ plotted as a function of initial amplitude $E_0$ using $p_{th}=0.3$ and $\delta=0.005$ (upper panel). For the lower panel, the damping $\Gamma_H$ is plotted as a function of $p_{th}$, for $E_0=1$ and $\delta=0.005$. The triangles show the period-time $T_p$ for the same parameters as used for the damping, as read off on the right-hand side axis.}
    \label{gamma_vs_init}
\end{figure}

Using the same background distribution as in the previous section, see \cref{background}, we study the dependence of wave damping on the initial temperature and wave amplitude.  The scaling of $\Gamma_H$ with the initial values of the plasma is displayed in \cref{gamma_vs_init}. In the upper panel, using the dots, we show the dependence of $\Gamma_H$ on the initial electric field amplitude. Somewhat surprisingly, perhaps, we see that the loss rate decreases with amplitude. To understand this, we note that the loss rate is normalized against the initial wave energy, and computed for a fixed unit time. Examining the expression for the loss rate \cref{Wloss} in the low-temperature limit, after a little algebra, it can be seen that the expression tends to scale as the energy density ($\propto E^2$) (compare the low-temperature result \cref{energy-cold-eq}), suggesting that the relative loss rate $\Gamma_H$, normalized against the initial energy, should be more or less independent of the wave amplitude. However, what has not been accounted for by such a simple consideration,  is the change in the temporal waveform. The sinusoidal profile for low amplitude has an average $\langle E^2(t)\rangle = (1/2)E_0^2$ over a wave period, whereas the average for a perfect sawtooth profile, as in the extreme nonlinear regime, is $\langle E^2(t)\rangle = (1/3)E_0^2$. The drop in loss rate seen in the upper panel of \cref{gamma_vs_init} is slightly lower, where the deviation can be considered as a thermal correction to the low-temperature result.    

However, it should be noted that while the relative loss rate per unit time drops (as displayed), the absolute loss rate per wave period actually grows. The reason is the increase in period time with amplitude.  
The right-hand vertical axis of the upper panel shows the simultaneous change of the period time of the oscillation using triangles. Due to the normalization, in the absence of relativistic effects, the period time would be $T_p=2\pi$, but for the larger amplitudes, we can note a substantial relativistic increase in the period time. In the upper panel, we have used $p_{th}=0.3$ and $\delta=0.005$. We note that for $E_n<<1$, we have $\Gamma \approx 0.03$. Here the relativistic gamma factors of the particles are close to unity.  For the larger electric fields, recalling that the drift momentum scales as $P\propto E^2$, we have strong relativistic effects. As $T_p\propto 1/\sqrt{\langle \gamma \rangle}$ (due to $\epsilon$ in the denominator of \cref{Amper'slaw2}) it is not surprising that the period time $T_p$ deviates from the non-relativistic result by almost an order of magnitude for large amplitudes. 

In the lower panel of \cref{gamma_vs_init}, the dots show the dependence of $\Gamma_H$ on the thermal momentum $p_{th}$, for initial electric field $E_0=1$ and $\delta=0.005$. The reason for the decay in $\Gamma_H$ with $p_{th}$ is that for a higher temperature, a gradually higher fraction of the energy loss comes from cooling the distribution, rather than from wave damping. The triangles again show the simultaneous variation of the period time $T_p$ on the vertical right-hand axis.  As is well-known, both a relativistic temperature and a relativistic wave amplitude increase the period time of the plasma oscillation, as confirmed by the smooth increase of the period time with thermal momentum.

\begin{figure}
    \centering
    \includegraphics[width=9cm,height=9cm]{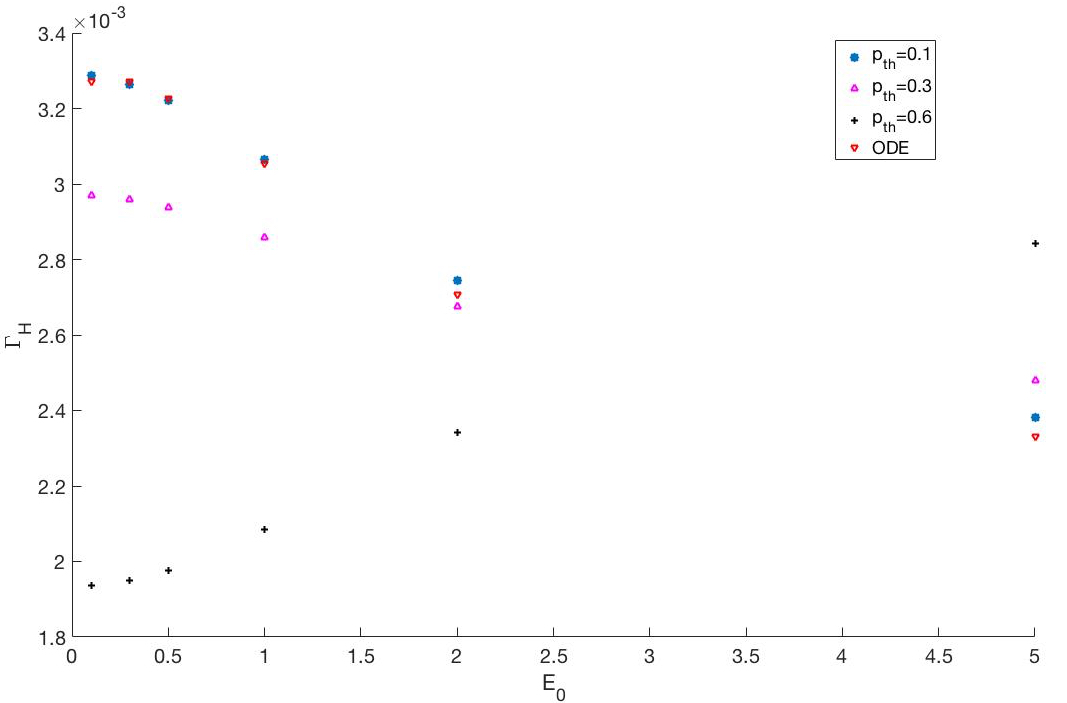}
    \caption{ $\Gamma_c$ as a function of the initial amplitude $E_0$ for $\delta=0.005$ and $p_{th}=(0.1,0.3,0.6)$ and the low-temperature solution "ODE".}
    \label{Gamma_vs_init}
\end{figure}

Finally, in \cref{Gamma_vs_init}, the dependence of the energy loss rate $\Gamma_H$ on initial wave amplitude $E_0$ is shown in the cold limit, as well as for three different nonzero values of the thermal momentum. For $p_{th}=0.1$, we see that for such a low thermal momentum, the loss rate deviates from the cold ($T=0$) result only very slightly, with the different data points more or less overlapping. For larger thermal momentum, and a small electric field amplitude, the loss rate decreases with $p_{th}$(as expected from the second term of \cref{Wloss}). For low thermal momentum, the curves show a consistent decline of $\Gamma_H$ with initial amplitude. This decline with amplitude is consistent with a transition from $\langle E^2(t)\rangle = (1/2)E_0^2$ for a low-amplitude sinusoidal profile, to $\langle E^2(t)\rangle = (1/3)E_0^2$ for a strongly nonlinear sawtooth profile. However, for a high thermal momentum, $p_{th}=0.6$, the scaling is radically different. Instead of the damping decreasing with initial amplitude, there is a steady increase of $\Gamma_h$ with $E_0$. A key factor here is that in this regime, much of the energy loss in \cref{temp5} and \cref{Wloss} comes from the term $\propto p_{\perp}^2 E^2$, rather than the terms $\propto \partial E/\partial t$. For a modest amplitude, the terms in \cref{temp5} and \cref{Wloss} $\propto p_{\perp}^2 E^2$, describing the temperature loss and the total energy loss, respectively, cancel to a good approximation. Thus the energy loss is mainly due to cooling, and the difference between the terms contributes very little to the wave-damping rate. However, in the strong field regime, this is no longer true, and much of the wave damping comes from the two latter terms of the radiation reaction in \cref{RR1}. 

With the two sources of energy loss, wave damping and cooling studied separately, we would like to compare the magnitude of the different sources to the high-frequency radiation energy. For this purpose, we define the ratio $R$ of background cooling relative to the total radiated energy, given by
\begin{equation}
R=\frac{\delta W_c}{\delta{W_{tot}}}=\frac{\delta W_c}{\delta{W_c}+\delta W_w}   
\end{equation}
where the different contributions are computed after a period time $T_p$ with the aid of \cref{Wloss} and \cref{temp4}. 

\begin{figure}
    \centering
    \includegraphics[width=0.5\textwidth ]{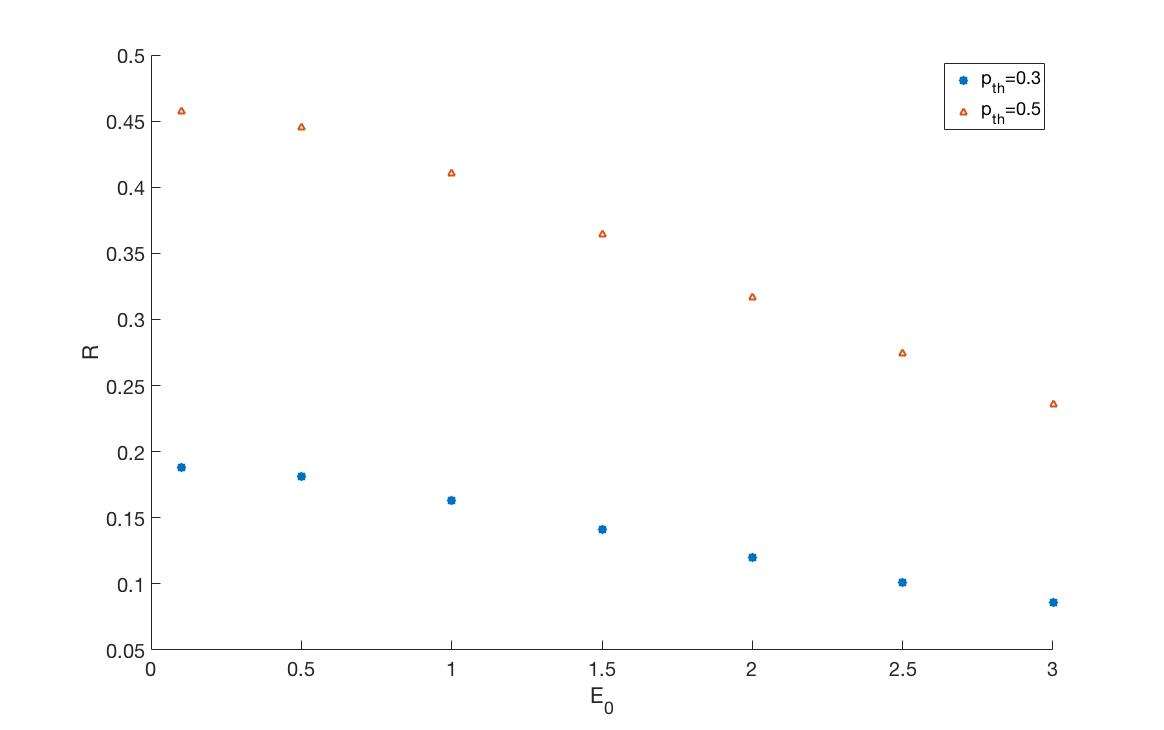}
    \caption{The ratio R (cooling energy loss over total energy loss) as a function of initial field amplitude $E_0$ using $\delta=0.0005$ and $p_{th}={0.3,0.5}$.}
    \label{R_vs_E}
\end{figure}

The ratio R is displayed in \cref{R_vs_E} as a function of initial amplitude for two different values of the thermal momentum. We see that the cooling as a source of energy for the high-frequency emission becomes more prominent for a higher temperature, as expected. Moreover, it can be seen that the relative contribution from cooling is decaying slowly with the initial amplitude. This may seem to contradict previous findings, as it has been found that the cooling grew strongly with $E_0$ (lower panel of \cref{Cooling}), whereas the wave damping either decayed (lower temperature) or grew moderately (higher temperature) with $E_0$ (see, \cref{Gamma_vs_init}). However, it is natural to normalize the wave energy loss against the initial wave energy, to get a wave damping rate, whereas the cooling energy loss naturally is normalized against the initial temperature, to get a cooling rate. While the cooling rate displayed in the lower panel of \cref{Cooling} shows a strong increase with $E_0$, it is still slightly slower than the growth of the wave energy density with initial amplitude ($\propto E_0^2$). Hence the scaling displayed in previous figures is indeed consistent with a slight decrease of $R$ with $E_0$, as shown in \cref{R_vs_E}.  

\section{Discussion and Conclusion}
In the present work, we have studied the influence of radiation reaction on Langmuir waves, by extending the relativistic Vlasov equation to include the Landau-Lifshitz form of the particle self-interaction. As expected, it is found that the Langmuir waves are damped, with an overall scaling proportional to $\delta=\omega_p r_e/c$, i.e. proportional to the square root of the electron number density. The reason the damping rate is not linear in the number of particles radiating is that the temporal scale s normalized against the plasma frequency, which in itself scales as $\sqrt{n_0}$

An energy conservation law containing the loss rate has been deduced, together with an equation for the temperature evolution,  which helps facilitate the damping rate as a function of initial temperature, initial amplitude, and wave number. Somewhat surprisingly, the normalized energy loss rate (to the initial wave energy) shows a slow decay with initial amplitude, unless the temperature is high, in which case the damping rate tends to grow with amplitude. The decrease in wave damping for low temperatures is a direct consequence of the transition from a sinusoidal wave profile for low amplitudes, to a saw-tooth profile in the large amplitude relativistic regime. 

In addition to the damping rate of Langmuir waves, it is found that the kinetic energy of the background distribution diminishes during the evolution. The relative cooling rate (normalized against initial temperature) has a modest dependence on the initial temperature, with a slightly higher relative cooling rate for a higher initial temperature. However, there is a relatively strong dependence of the cooling rate on the wave amplitude. In particular, a stronger amplitude gives a cooling almost, but not quite, scaling as $\propto E_0^2$. This behavior is quantified in the ratio $R$, describing how much of the emitted high-frequency radiation comes from the background kinetic energy, as opposed to the fraction that comes from wave damping. here we see only a weak dependence on $E_0$. Since the wave energy loss is $\propto E_0^2$ in the crudest of approximations, the same goes for the cooling of the background distribution. 

In the present work, we have limited ourselves to electrostatic waves, but it is of much interest to also cover electromagnetic waves. A principal difference is that for electrostatic waves, the maximum electric field amplitude is limited by the electron number density, whereas for electromagnetic waves, there is no direct upper bound. Moreover, in an electrostatic geometry, the particle acceleration tends to be approximately parallel to the velocity, which limits the magnitude of the radiation reaction. 

As a result, radiation reaction in an electromagnetic context may take place in a regime, where the radiation reaction is close to comparable with the Lorentz force, in contrast to the case studied here. In such a regime, models extending the Landau-Lifshitz force, such as e.g. the regulated Lorentz-Abraham-Dirac theory  \cite{Robin-RR, Greger-RR} and/or the quantum corrected theory \cite{Q-korr}may be of interest, as well as further quantum extensions, see e.g. Refs. \cite{quantum-1, quantum-2}

 \bibliography{References}

\end{document}